\documentclass[twocolumn,aps,prl,superscriptaddress,showpacs,amsmath,amssymb,floats]{revtex4}

\usepackage{amsmath}
\usepackage{amssymb}
\usepackage{graphicx}
\usepackage{CJK}
\usepackage{keyval}
\usepackage{dcolumn}
\usepackage{bm}
\usepackage{color}
\usepackage{txfonts}
\usepackage[]{sidecap}

\begin{document}

\begin{CJK*}{UTF8}{gbsn}

\title{Anisotropic but nodeless superconducting gap in the presence of spin
density wave in iron-pnictide superconductor NaFe$_{1-x}$Co$_{x}$As}

\author{Q. Q. Ge (\CJKfamily{gbsn}葛青亲)}

\author{Z. R. Ye (叶子荣)}

\author{M. Xu (徐敏)}

\author{Y. Zhang  (张焱)}

\author{J. Jiang (姜娟)}

\author{B. P. Xie (谢斌平)}

\affiliation{State Key Laboratory of Surface Physics, Department of Physics, and
Advanced Materials Laboratory, Fudan University, Shanghai 200433,
People's Republic of China}

\author{Y. Song (宋宇)}

\affiliation{Department of Physics and Astronomy, The University of Tennessee,
Knoxville, Tennessee 37996-1200, USA}

\author{C. L. Zhang (张承林)}

\affiliation{Department of Physics and Astronomy, The University of Tennessee,
Knoxville, Tennessee 37996-1200, USA}

\author{Pengcheng Dai (戴鹏程)}

\affiliation{Department of Physics and Astronomy, The University of Tennessee,
Knoxville, Tennessee 37996-1200, USA}

\affiliation{Beijing National Laboratory for Condensed Matter Physics, Institute
of Physics, Chinese Academy of Sciences, Beijing 100190, China}

\author{D. L. Feng (封东来)}

\email{dlfeng@fudan.edu.cn}

\affiliation{State Key Laboratory of Surface Physics, Department of Physics, and
Advanced Materials Laboratory, Fudan University, Shanghai 200433,
People's Republic of China}

\date{\today}
\begin{abstract}
The coexisting regime of spin density wave (SDW) and superconductivity
in iron pnictides represents a novel ground state. We have performed
high resolution angle-resolved photoemission measurements on NaFe$_{1-x}$Co$_{x}$As
($x=0.0175$) in this regime and revealed its distinctive electronic structure, which provides some microscopic understandings of its behavior.
The  SDW signature and the superconducting gap are observed
on the same bands, illustrating the intrinsic nature of the coexistence.
However, because the SDW and superconductivity are manifested in different
parts of the band structure, their competition is  non-exclusive. Particularly, we found that the gap distribution is
anisotropic and nodeless, in contrast to the isotropic superconducting
gap observed in an SDW-free NaFe$_{1-x}$Co$_{x}$As (x=0.045), which puts strong constraints on theory.

\end{abstract}


\maketitle
\end{CJK*}

Most unconventional superconductors appear in the vicinity of a certain
magnetically ordered phase \cite{RevUnconventionalSC}. Magnetism
is suggested to play a critical role in the pairing mechanisms of
the cuprates \cite{Cuprate}, heavy Fermion superconductors \cite{Cuprate,CeCu2Si2},
and even organic superconductors \cite{OrganicSC}. For iron-pnictide
superconductors, a spin density wave (SDW) phase appears next to the
superconducting (SC) phase \cite{HosonoLaOFeAs,XHChenSmFeAsO,cruz}, and
in some cases, they even coexist
\cite{NeutronSc1,NeutronSc&Theory,NeutronSc3,BaCoTrans,BaCoTrans2,ZhangSrK}, which gives a unique
SC ground state.
While the coexisting SDW and SC
phases may have significant impact on the SC mechanism \cite{NeutronSc&Theory}, much is not clear
about the subtle interacting nature between magnetism and
superconductivity \cite{hqluo}.
In fact, theories based on $s^{++}$ pairing symmetry suggest that
there must be nodes in the SC  gap in this regime \cite{theoryMazin}
and the coexisting SDW and SC phases cannot be microscopic \cite{NeutronSc&Theory}.
On the other hand, theories based on $s^{+-}$ pairing symmetry suggest
nodeless SC gap in the presence of weak magnetic order; moreover,
the coexistence may cause angular variation of the SC gap, and even give
rise to nodes in the limit of strong antiferromagnetic (AFM) ordering
\cite{theoryMazin,theoryChubukov}, as indicated in a thermal conductivity
study on Ba$_{1-x}$K$_{x}$Fe$_{2}$As$_{2}$ \cite{Node1}.

\begin{figure*}[t]
\includegraphics[width=17cm]{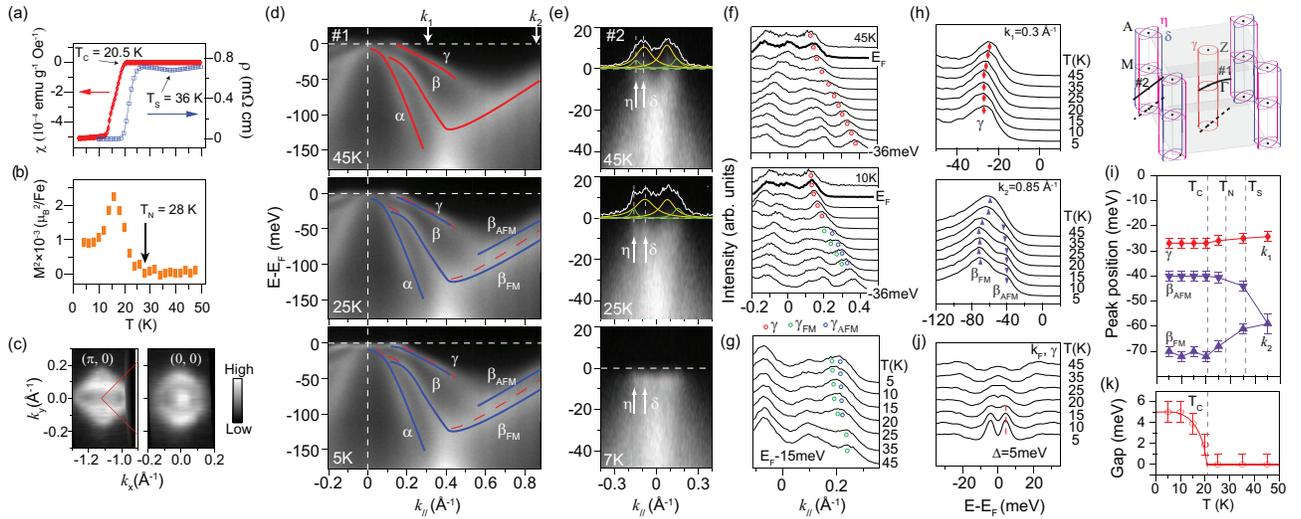} \caption{(color online)
(a) Magnetic susceptibility of NaFe$_{0.9825}$Co$_{0.0175}$As
single crystal taken at a magnetic field of 20~Oe in the zero field cool mode, and its resistivity
as a function of temperature.  (b) Temperature dependence of the magnetic order parameter at Q= (1, 0, 1.5) for NaFe$_{0.9825}$Co$_{0.0175}$As measured by neutron scattering. (c) Photoemission intensity map at the Fermi energy integrated over [$E_{F} - 5$ meV, $E_{F} + 5$ meV].  (d) The band structure of
NaFe$_{0.9825}$Co$_{0.0175}$As at 45, 25, and 5 K respectively along cut~\#1 across $\Gamma$ as indicated in the inset.
The dashed lines in the lower panels are the band dispersion at 45~K for comparison purpose. (e) Temperature dependence of the band structure around the zone
corner along cut~\#2 as indicated in the inset. The MDCs (momentum distribution curves) at E$_{F}$ are plotted on
the 25 and 45 K data. Each MDC was fitted to four Lorentzians (overlaid yellow and green lines). (f) MDCs near the zone center at 45 and 10~K. (g) Temperature dependence of the MDC at $E_F-15$ meV near the zone center. The positions of the $\gamma$ band are marked in panels (f) and (g).
(h) Temperature dependence of the EDCs (energy distribution curves) at selected
momenta: $k{}_{1}$= 0.3 $\AA^{-1}$, and $k{}_{2}$= 0.85 $\AA^{-1}$
respectively as marked in panel (d). Due to the broad lineshape,  $\gamma_{AFM}$ and $\gamma_{FM}$ are not resolved, but the shift of the overall features is obvious. (i) The temperature dependence
of the  peak positions in panel (h). (j) The temperature dependence
of the symmetrized EDCs measured at the $k_{F}$ of  $\gamma$. (k) The temperature dependence of the superconducting (SC) gap of $\gamma$. The gap size is estimated through an empirical fit as described in detail
in Ref.~\cite{yanBaPNode}.  The inset on the top right corner shows the Fermi surface of NaFe$_{0.9825}$Co$_{0.0175}$As. The two solid lines mark cut~\#1 and cut~\#2 along which the
data in panels (d) and (e) are located, respectively. The two dashed lines on the bottom plane are their projections.
The photoemission data in panels (c) and (e) were acquired in-house, and others were collected at SSRL. }
\label{bands}
\end{figure*}

The coexistence of SDW and superconductivity in various iron pnictides
has been illustrated by neutron scattering \cite{NeutronSc1,NeutronSc&Theory,NeutronSc3,BaCoTrans,BaCoTrans2}, nuclear magnetic resonance \cite{NMRBaCoK,NMRBaCo},
and angle-resolved photoemission
spectroscopy (ARPES) experiments \cite{ZhangSrK}. Recent scanning tunneling
microscope (STM) studies show the real-space coexistence and competition
of SDW and superconductivity in NaFe$_{1-x}$Co$_{x}$As \cite{Yayucoexist,Yayucoexist2}
. However so far, little is known regarding the electronic structure
of the coexisting phase in the momentum space, such as its SC gap
distribution, and how the two orders coexist and compete on the same
electronic structure. In this paper, we report ARPES studies on NaFe$_{0.9825}$Co$_{0.0175}$As
in this coexisting regime. The band structure reconstruction corresponding
to the SDW formation and the SC gap could be observed on the same bands, which
provides a direct evidence for the intrinsic coexistence of the two
orders. We found that SDW formation does not cause much depletion of the states
near the Fermi energy ($E_{F}$), therefore, it allows the superconductivity
to occur. Moreover, the SC gap distribution
is found nodeless on all Fermi surface sheets: it is isotropic on
the hole pocket, but it is highly anisotropic on the electron pockets.
Our results reveal the distinct electronic properties of the coexisting
phase and provide explicit constraints on theory.

High-quality NaFe$_{0.9825}$Co$_{0.0175}$As single crystals were
synthesized by the self-flux method described elsewhere \cite{heNa}.
The SC transition temperature ($T_{c}$) is determined by the magnetic
susceptibility measurements with a SQUID magnetometer [Fig.~\ref{bands}(a)],
which shows an onset drop at 20.5~K. Resistivity measured by PPMS
indicates zero resistivity below 18~K, and a structural transition
at $T_{S}=36$ K. Our neutron scattering data show that the SDW transition temperature ($T_N$) is 28~K [Fig.~\ref{bands}(b)]. ARPES data were taken with various photon energies
in circular polarization at the 1-Cubed beamline of BESSY II, other photoemission
measurements were performed either with 21~eV photons at beamline
5-4 of the Stanford Synchrotron Radiation Laboratory (SSRL), or with
randomly polarized 21.2~eV light from an in-house SPECS UVLS helium
discharging lamp at Fudan University. All the data were taken with
SCIENTA R4000 electron analyzers; the overall resolution is set to
6~meV or better and the typical angular resolution is $0.3^{\circ}$.
The samples were cleaved {\it in situ}, and measured under ultra-high
vacuum, so that the aging effects are negligible in the data.

The general electronic structure of NaFe$_{0.9825}$Co$_{0.0175}$As is rather
similar to the well studied NaFeAs \cite{heNajpcs,yanNa,heNa}. Figure
\ref{bands}(c) shows the photoemission intensity map near $E_{F}$
taken at 7~K with 21.2~eV photons. There are a hole pocket and a small
patch-like feature point around $\Gamma$ $(0,0)$, and two orthogonal elliptical
pockets around the zone corner. The photoemission intensity along cut \#1 across $\Gamma$ is
plotted in Fig.~\ref{bands}(d), where three bands, $\alpha$, $\beta$
and $\gamma$ could be resolved, but only $\gamma$ crosses $E_{F}$
and gives the hole Fermi surface. The band top of $\alpha$ is just
below $E_{F}$, and contributes to the small patch in the zone center.
Figure \ref{bands}(e) plots the photoemission intensities at the zone
corner, where two electron-like bands, $\delta$ and $\eta$, could
be observed. As previous photon energy dependent study has revealed
the negligible $k{}_{z}$ dispersion of NaFeAs \cite{heNajpcs}, the
overall Fermi surface topology of NaFe$_{0.9825}$Co$_{0.0175}$As
is summarized in the inset on the top right corner of Fig.~\ref{bands}.

The signature of SDW on the electronic structure has been extensively
studied before \cite{heNa,yanNa,MingYiNa,LXyang}, which is mainly
manifested as a remarkable band reconstruction. As shown in Fig.~\ref{bands}(d),
$\beta$ shifts significantly with decreased temperature.  To  illustrate
the subtle band reconstruction of $\gamma$, Fig.~\ref{bands}(f) plots the momentum distribution
curves (MDCs) near the Fermi crossing of $\gamma$
at several binding energies near $E_{F}$ at 45 and 10~K, and Fig.~\ref{bands}(g) plots the MDC at $E_F-15$ meV as a function of temperature. It is clear that $\gamma$ first shifts in one direction due to the
SDW \cite{yanNa}, and then splits into two at low temperatures.
Our recent ARPES study on the mechanically detwinned NaFeAs has shown that the
$\beta$ and $\gamma$ bands disperse differently along the ferromagnetic (FM) and
AFM directions, which gives an appearance of band
splitting in the twinned sample here as noted by the subscripts in Fig.~\ref{bands} \cite{yanNa}. Similar reconstruction effects can be observed in the energy distribution curves (EDCs) as well in Fig.~\ref{bands}(h). As shown by the temperature dependence of the EDC
peak positions summarized in Fig.~\ref{bands}(i), the electronic structure reconstruction
occurs above the structural transition due to the fluctuations of
the SDW and electronic structure nematicity \cite{yanNa,MatsudaNematic}. It evolves smoothly across the structural  and Neel transitions, and saturates below 20 K, with the separation of $\beta_{AFM}$ and
$\beta_{FM}$ reaching 32~meV and the shift of $\gamma$ reaching
3~meV. The reconstruction of $\delta$ and $\eta$ is subtle, nevertheless
in Fig.~\ref{bands}(e), their features in the MDCs at $E_{F}$ clearly show finite shifts as well \cite{yanNa}.
On the other hand, SC gap opens just below $T_c$, as illustrated
by the symmetrized EDCs of the $\gamma$ band with respect to $E_{F}$
in Fig.~\ref{bands}(j) and the fitted SC gap in Fig.~\ref{bands}(k). The fact that the
signatures of both the superconductivity and SDW emerge in the same
band structure confirms their intrinsic coexistence. Furthermore, the band reconstruction due to SDW mainly occurs over a large energy and momentum scales for $\beta$ below $E_F$, and
it leaves the states on all the Fermi surfaces largely intact in this doping regime, therefore
superconductivity could occur in the presence of SDW here.

\begin{figure}[t!]
\includegraphics[width=8.5cm]{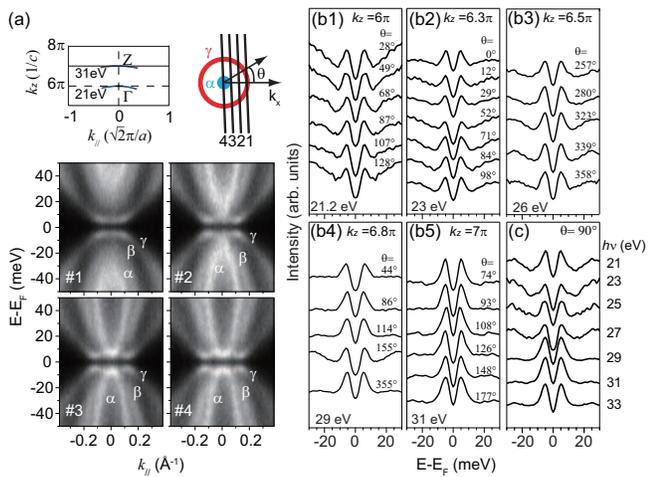} \caption{(color online) (a) The photoemission intensities taken with 21.2~eV
photons at 7~K near the zone center as shown by cuts $\#1-\#4$ in
the inset. The inset to the left shows the momentum cuts sampled by
the 21 and 31~eV photons in the $k_{x}-k_{z}$ cross-section of the
extended Brillouin zone. (b) The symmetrized spectra at the marked
polar angles on the $\gamma$ Fermi surface measured at five typical
$k{}_{z}$ values with (b1) 21.2, (b2) 23, (b3) 26, (b4) 29, and (b5)
31~eV photons. (c) $k{}_{z}$ dependence of the symmetrized spectra measured on the $\gamma$ Fermi
surface of another sample at $\theta=90^{\circ}$. The 21.2~eV data were collected at 7~K with a helium
lamp, while the others were collected at 1~K at BESSY.  }
\label{gap1}
\end{figure}

\begin{SCfigure*}[][b!]
\includegraphics[width=12cm]{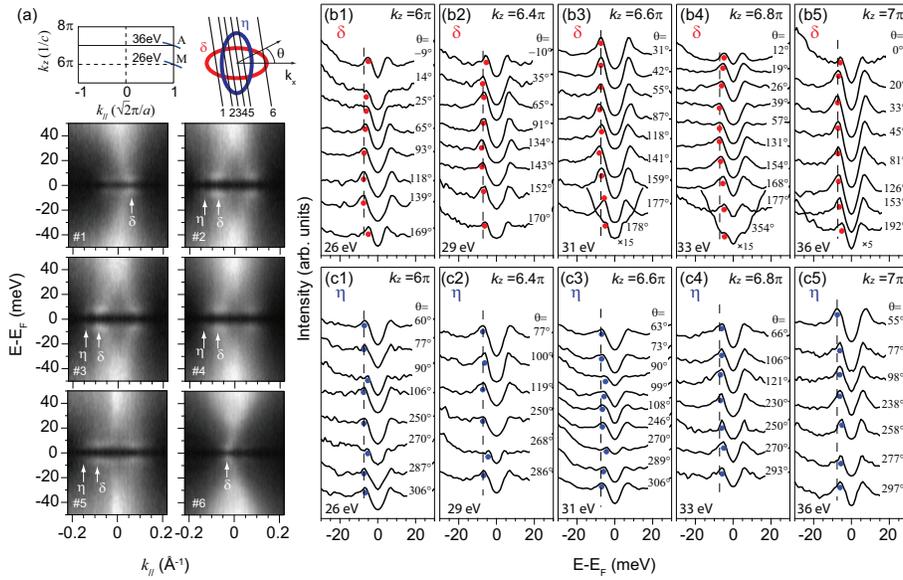} \caption{(color online) (a) The photoemission intensities taken with 26~eV
photons at 1~K near the zone corner as shown by cuts $\#1-\#6$ in the top-right inset. The top-left inset shows the momentum cuts sampled by
the 26 and 36~eV photons in the $k_{x}-k_{z}$ cross-section of the
Brillouin zone. (b) The symmetrized spectra at the marked polar angles
on the $\delta$ Fermi surface measured at five typical $k{}_{z}$'s with (b1) 26, (b2) 29, (b3) 31, (b4) 33, and (b5) 36~eV photons.
The dashed lines here are guides to the eyes for the variation of
the SC gaps. (c1)-(c5) Same as in panels (b1)-(b5) but on the $\eta$
electron Fermi surface.   All  data were
collected at 1 K in BESSY-II. Note that the bottom curves in panels (b3),
(b4), and (b5) are magnified vertically.\\
\\
}
\label{gap2}
\end{SCfigure*}

\begin{figure}[t!]
\includegraphics[clip,width=8.7cm]{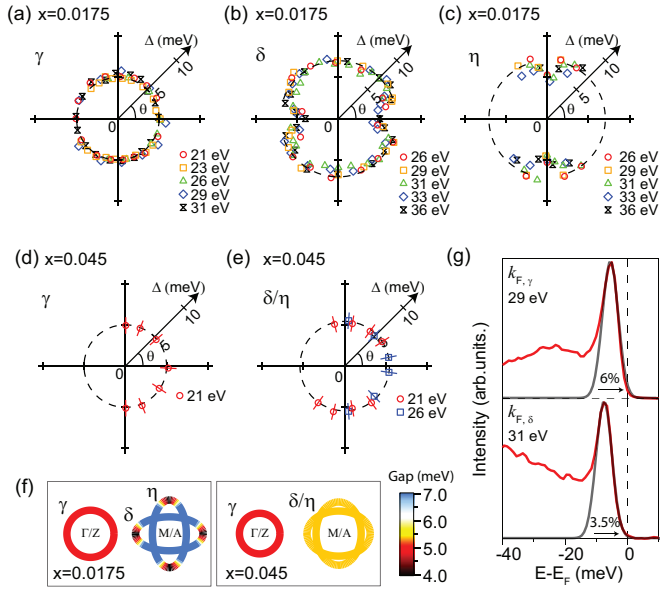} \caption{(color online) Polar plots of the SC gap for the (a) $\gamma$, (b)
$\delta$ and (c) $\eta$ Fermi surfaces of NaFe$_{0.9825}$Co$_{0.0175}$As,
respectively. The polar angle $\theta$ follows the same definition
in Figs.~\ref{gap1} and \ref{gap2}. The error bar for the gaps is
$\pm$1~meV based on the fitting. Polar plots of the SC gap of NaFe$_{0.955}$Co$_{0.045}$As for the (d) $\gamma$,  and (e) $\delta/\eta$
Fermi surfaces respectively. (f) False-color plots of the gap distribution
on the Fermi surfaces of NaFe$_{0.9825}$Co$_{0.0175}$As and NaFe$_{0.955}$Co$_{0.045}$As. (g) The typical spectra at the Fermi crossings of the $\gamma$
and $\delta$ bands taken at 1 K in BESSY-II.  The intensity ratio of
the residual spectral weight at $E_F$ is referred to the coherence peak height. Two  Gaussians  with 6~meV full-width-half-maximum are overlaid. }

\label{sum}
\end{figure}

The SC gap is mapped out extensively over the entire Brillouin zone.
Figure \ref{gap1}(a) shows the symmetrized photoemission intensity
along four momentum cuts across the $\gamma$ hole Fermi surface in
the $k_{z}=6\pi$ plane. The suppression of the spectral weight around
$E_{F}$ indicates the opening of the SC gap. In Fig.~\ref{gap1}(b1),
the symmetrized EDCs along the $\gamma$ pocket clearly show sharp
coherent peaks, and  SC gaps of similar amplitude. Data from other $k_{z}$ planes
in Figs.~\ref{gap1}(b2)-\ref{gap1}(b5), and data from another sample
taken with more photon energies in Fig.~\ref{gap1}(c) show that
the gap is isotropically 5~meV on the $\gamma$ pocket, as also summarized in Fig.~\ref{sum}(a).

Now we turn to the SC gap on the electron Fermi surfaces around the
zone corner. Figure \ref{gap2}(a) shows symmetrized photoemission
intensity for six momentum cuts across the $\delta$/$\eta$ pockets
in the $k{}_{z}=6\pi$ plane, where the SC gap opens on both Fermi
surfaces. Collecting the symmetrized EDCs at various $k_{F}$'s along
the $\delta$ pocket, Fig.~\ref{gap2}(b1) demonstrates an anisotropic
gap distribution, where the gap is about 7~meV in the flat part of
the ellipse, and significantly drops to 4~meV near $\theta=0^{\circ},180^{\circ}$.
Moreover, such a behavior is observed for all five sampled $k{}_{z}$'s
as shown in Figs.~\ref{gap2}(b1)-\ref{gap2}(b5). Similarly, such
an anisotropic gap distribution is observed for $\eta$ but rotated
by $90^{\circ}$ [Figs.~\ref{gap2}(c1)-\ref{gap2}(c5)]. The
weak $k{}_{z}$ dependence is further illustrated with more data taken at  $k_z=5.5\pi$, $6.3\pi$, and $6.5\pi$ with 21, 28, and 30~eV photons respectively in the supplementary material [Fig.~S1].

The gap distribution of NaFe$_{0.9825}$Co$_{0.0175}$As is summarized
in Figs.~\ref{sum}(a)-\ref{sum}(c). The gaps along the $\gamma$
hole Fermi surface show isotropic distribution, while the gaps on
the $\delta$ and $\eta$ pockets vary significantly from 4 to 7~meV.
As a comparison, Figures \ref{sum}(d)-\ref{sum}(e) show the isotropic in-plane gap distribution on individual Fermi
surfaces for an SDW-free NaFe$_{0.955}$Co$_{0.045}$As sample ($T_{c}=20$~K),  which are retrieved from the symmetrized EDCs provided in the supplementary material [Fig.~S2]. The gap is about 5~meV on the hole pocket,
and 5.4~meV on the electron pockets.  Such an isotropic in-plane  gap  distribution has been observed before in NaFe$_{0.95}$Co$_{0.05}$As as well \cite{SCWangNaCo}.  Furthermore, Fig.~\ref{sum}(f)
compares both the Fermi surfaces and the SC gap distributions of NaFe$_{0.9825}$Co$_{0.0175}$As
and NaFe$_{0.955}$Co$_{0.045}$As. The hole pocket of NaFe$_{0.955}$Co$_{0.045}$As
is slightly smaller as expected from cobalt doping, and the ellipticity
of its electron pockets is smaller as well.

So far in ARPES experiments, the in-plane anisotropy of SC gap has  been observed only for LiFeAs \cite{LiFeAs1,LiFeAs2},
Fe(Te,Se) \cite{InplaneAnisotropyFeTeSe},  KFe$_{2}$As$_{2}$ \cite{InplaneAnisotropyKFe2As2}, and  Ba$_{1-x}$K$_x$Fe$_2$As$_2$ \cite{InplaneBaK} among all the iron-based superconductors,
but none of them is in the coexisting regime. The small gap anisotropy on one of the hole pockets of Ba$_{1-x}$K$_x$Fe$_2$As$_2$ is within the experimental error that less than 0.6~meV difference over the 9$\sim$10~meV gap amplitude is observed \cite{InplaneBaK}.  The moderately anisotropic gap on a hole Fermi
surface of LiFeAs might be a mere consequence of the
Fermi surface topology, since it is qualitatively consistent with the gap function $\triangle(k)=\triangle_{0}cosk_{x}cosk_{y}$
predicted based on the $s{}^{+-}$pairing symmetry \cite{LiFeAs1,LiFeAs2}. For NaFe$_{0.9825}$Co$_{0.0175}$As, the large ellipticity gives a variation of $|cosk_{x}cosk_{y}|$ from
$\sim$ 0.98 in the flat region to $\sim$ 0.91 on the tip, which
could not explain the over 40\% change of the gap based on the Fermi
surface topology.   We note that an anisotropic gap distribution around the zone corner has also been revealed in LiFeAs, which deviates from the canonical $s^{+-}$-wave gap function and was explained in terms of the band hybridization \cite{LiFeAs2}. Consistently, the diviation there is most prominent around $\theta=45 ^\circ$ where the hybridization is the strongest. However, the anisotropic behavior in NaFe$_{0.9825}$Co$_{0.0175}$As deviates the gap function remarkably around $\theta=0$ and $90 ^\circ$, which is away from Fermi surface region of mixed orbital character.
For Fe(Te,Se), the anisotropy of the SC gap on the hole pocket was
suggested to be a consequence of sizable second-nearest-neighbor
interactions, while the anisotropic and nodal gap on a hole pocket of KFe$_{2}$As$_{2}$ may be related to strong intra-pocket scattering \cite{nodeDHLee}, or specific orbital characters near Z \cite{yanBaPNode}. Alternatively, the angular variation in the $d_{xy}$
orbital content of the $\gamma$ Fermi surface was predicted to cause
anisotropic gap distribution on the electron pockets \cite{nodeDHLee}.
However, since NaFe$_{0.9825}$Co$_{0.0175}$As and
NaFe$_{0.955}$Co$_{0.045}$As have similar Fermi surface, orbital characters and
interaction parameters, NaFe$_{0.955}$Co$_{0.045}$As would have
exhibited anisotropic gap if these had been the causes here. Therefore,
the highly anisotropic gap distribution on the electron pockets of
NaFe$_{0.9825}$Co$_{0.0175}$As is most likely a direct consequence
of the coexisting SDW.


Theories based on the $s{}^{+-}$ paring symmetry have suggested the
nodeless and anisotropic gap distribution in the presence of weak
SDW \cite{theoryMazin,theoryChubukov}. Consistently, compared with
NaFeAs \cite{yanNa},  much weaker SDW order is present in NaFe$_{0.9825}$Co$_{0.0175}$As: the band folding due to the SDW order is negligible, and no SDW gap induced by the hybridization
with the folded bands is observed here.
In a recent theoretical study, it was predicted that even weak SDW order will cause appreciable gap anisotropy  \cite{theoryChubukov}. Particularly, it was found that the gap at the tip region  of the electron Fermi surface is smaller than that at the flat region, in good agreement with our observation.
Futhermore, the observed nodeless SC gap disallows the paring mechanism based on the $s^{++}$ pairing symmetry
that predicts SC gap nodes in the SDW state \cite{NeutronSc&Theory,theoryMazin}.

The prominent band reconstruction of $\beta$ observed here with
a 32 meV separation between the dispersions along the AFM
and FM directions is smaller than the 46 meV observed in NaFeAs \cite{heNa}.
Such a band reconstruction energy scale is distinct at a specific doping, and is correlated with the SDW transition temperature as observed in Sr$_{1-x}$K$_x$Fe$_2$As$_2$ \cite{ZhangSrK}. Therefore, the sharp band dispersion with a single
set of band reconstruction energy scale, plus the resolution limited
width of the superconducting coherent peak [Fig.~\ref{sum}(g)], highlight the homogeneous nature of the electronic state in the momentum space. Moreover, although the shielding fraction of the bulk sample is 75\%  based on our susceptibility measurements, the ARPES data are taken on a small region ($0.05~mm\times 0.2~mm$) of the cleaved surface. As shown in Fig.~\ref{sum}(g), the photoemission intensity at $E_{F}$ in the superconducting state is negligible, which suggests the absence of non-superconducting region. That is, there is no phase separation of superconducting regions and non-superconducting SDW regions in the coexisting phase. Our results thus rule out the appearance of macroscopic phase separation and further support the intrinsic coexistence. These are consistent with a recent STM study on the coexisting phase of NaFe$_{1-x}$Co$_{x}$As
(x=0.014) \cite{Yayucoexist2}, where the coexistence was found to
occur microscopically in an anti-correlated but non-exclusive way
between the two orders. Such a non-exclusive coexistence can be understood based on our observation of the indirect competition between SDW and superconductivity in the electronic structure. Note that, the energy scales observed in STM for both the ``SDW gap'' feature ($\sim$ 17~meV, and it should be a momentum-integrated effect of the band reconstruction) and SC
coherence peak ($\sim$ 5~meV) are quite independent of space. This is further
consistent with the single set of SDW/SC energy scales observed here by ARPES.

Our neutron scattering data on the same sample reveals that
static antiferromagnetic long-range order coexists with superconductivity, similar to the static antiferromagnetic order/superconductivity coexisting  BaFe$_{2-x}$Ni$_x$As$_2$ samples \cite{hqluo}. The intensity of the SDW diffraction peak decreases upon entering the SC state, suggesting a competition between the two orders [Fig.~\ref{bands}(b)]. The magnitude of the SDW order could be monitored directly from the energy scale of the band reconstruction. However, we did not observe any remarkable change of band reconstruction below $T_c$, which suggests that the competition between the two orders does not affect the magnitude of the local SDW order at the fast time scale of photoemission ($\sim 1~fs$). Alternatively, since the itinerant electrons near $E_F$ could play an important role in stabilizing the long-range SDW order \cite{mazin}, when the SC gap opens, the coherence of SDW order could be suppressed. Consequently, the enhanced fluctuation of the local SDW order  could be responsible for the observed suppression of the effective (or time-averaged) moment at the quasi-elastic neutron scattering time scale ($\gg 1~ps$) \cite{Mannella}.



To summarize, we have revealed detailed electronic structure in the
superconductivity/SDW coexisting regime of NaFe$_{1-x}$Co$_{x}$As
(x=0.0175), and signature in the momentum space for the intrinsic
microscopic coexistence. We found  that SDW does not cause a noticeable  depletion of the states at the Fermi energy, which allows the superconductivity to emerge. Therefore, it explains why the two orders could coexist in a non-exclusive way. Moreover, we show that the anisotropy
of the SC gap on the electron pockets is likely a distinct consequence of the coexisting SDW order, while the absence of gap node puts strong
constraints on the pairing symmetry in theory of iron-based superconductors.

We gratefully acknowledge the helpful discussions with Prof. J. P.
Hu and Prof. A. V. Chubukov, and the experimental support by Dr. D. H. Lu, Dr. M. Hashimoto
at SSRL, and Dr. E. Rienks at BESSY II. This work is supported in
part by the National Science Foundation of China and National Basic
Research Program of China (973 Program) under the grant Nos. 2012CB921400,
2011CB921802, 2011CBA00112. The single crystal growth efforts and neutron scattering work at University of Tennessee are supported by the US DOE, BES, through contract DE-FG02-05ER46202. SSRL is operated by the US DOE, BES, Divisions of Chemical Sciences and Material Sciences.

\end{document}